\RequirePackage{fix-cm}
\documentclass[aps,prd,amsmath,amssymb,superscriptaddress,twocolumn,nofootinbib,showpacs,preprintnumbers,notitlepage]{revtex4-1}      
\usepackage{graphicx}
\usepackage{amsmath}

\newcommand{\jlab}{Jefferson Lab, Newport News, Virginia 23606, USA}
\newcommand{\ut}{Institute for Theoretical Physics, T\"ubingen University, Auf der Morgenstelle 14, 72076 T\"ubingen, Germany}
\newcommand{\ur}{Institute for Theoretical Physics, Regensburg University, 93040 Regensburg, Germany}

\begin{document}
\preprint{JLAB-PHY-23-3773}

\title{Polarized proton structure in the resonance region}
\author{A.~N.~\surname{Hiller Blin}}
\affiliation{\ur}
\affiliation{\ut}
\author{V.~I.~\surname{Mokeev}}
\affiliation{\jlab}

\begin{abstract}
In view of the precise data available on inclusive polarized electron scattering off polarized proton targets in the nucleon resonance excitation region, we compare these results with the coherent sum of resonant contributions to the polarized structure function $g_1$ and virtual photon asymmetry $A_1$. To this goal, we employ the nucleon resonance electroexcitation amplitudes determined for photon virtualities $Q^2$ $<$ 5.0 GeV$^2$ from analyses of the CLAS data on exclusive electroproduction off protons in the resonance region. 
Most of the well established resonances of four star PDG status in the mass range up to 1.75~GeV are included.
We find that the resonance-like structures observed in the inclusive $g_1$ data are related to the resonant contributions in the entire range of photon virtuality $Q^2$ where the data on $g_1$ are available.
In the range of invariant mass of the final hadron system $W$ $>$ 1.5 GeV, the data on the asymmetry $A_1$ are well reproduced even when accounting for resonant contributions only, especially for the larger values of $Q^2$ and energies analysed. This observation offers an interesting hint to quark-hadron duality seen in polarized inclusive electron scattering observables.
\end{abstract}

\date{\today}

\maketitle

\section{Introduction}
\label{intro}
{ Inclusive} electron scattering off protons 
{and the exploration of its polarization observables offer} an essential means to obtaining insights about the {ground proton} structure~\cite{Jimenez-Delgado:2013sma,Gao:2017yyd,Ethier:2020way}. 
{The extension of these studies} to the resonance region will allow one to {understand the proton structure at large values of the fractional parton momentum $x$ in the resonance region and eventually to shed light on the strong interaction dynamics which underlies the transition from the strongly coupled to the perturbative QCD regimes,}  
as well as the associated characteristics of quark-hadron duality~\cite{Melnitchouk:2005zr,Lagerquist:2022tml,Bloom:1970xb,Osipenko:2003bu,Prok:2014ltt,Malace:2009kw,Christy:2007ve,Tvaskis:2016uxm,Liang:2004tj}.

There have been {impressive} advances in measuring inclusive scattering of polarized electron beams off polarized nucleon targets~\cite{E143:1998hbs,CLAS:2003rjt,Bosted:2006gp,CLAS:2006ozz,RSS:2006tbm,RSS:2008ceg,CLAS:2014qtg,CLAS:2017qga}, which open the path to duality studies in spin-dependent observables~\cite{Close:1972ah,Carlson:1998gf,Edelmann:1999yp,Melnitchouk:2005zr,Lagerquist:2022tml}. In order to improve the theory approaches describing the connection between resonances and scaling contributions, considerations need to be made about the role of the nonresonant background. While such a quantitative description from first principles is rather challenging, insight may be obtained from phenomenological analyses.

The experimental program exploring exclusive $\pi^+ n$, $\pi^0 p$, $\eta p$, and $\pi^+ \pi^-p$ electroproduction channels in the resonance region with the CLAS detector at Jefferson Lab has provided important new information on the $\gamma^* p N^*$ electrocouplings of most nucleon resonances in the mass range $W \leq 1.75$~GeV and { for} $Q^2 \leq 5$~GeV$^2$~\cite{Aznauryan:2011qj,Aznauryan:2009mx,Mokeev:2012vsa,Mokeev:2015lda,Park:2014yea,Carman:2020qmb,Mokeev:2020hhu,Burkert:2022ioj}.
These results allow one to quantitatively evaluate the coherent sum of resonant contributions to inclusive electron scattering observables, using parameters of the individual nucleon resonances extracted from data. 

In our previous works~\cite{Blin:2019fre,Blin:2021twt,HillerBlin:2022ltm}, we confronted polarized and unpolarized inclusive electron-scattering data with the computation of resonant contributions in the resonance region.
In the present work, we include updated data on the polarized structure function $g_1$~\cite{CLAS:2017qga} and the virtual photon asymmetry $A_1$~\cite{E143:1998hbs,CLAS:2003rjt,Bosted:2006gp,CLAS:2006ozz,RSS:2006tbm,RSS:2008ceg,CLAS:2014qtg,CLAS:2017qga}. In particular, the latter 
{ have extended} the coverage in $Q^2$ and $W$ {in comparison with} the data {analyzed} in our previous work, therefore permitting more insightful conclusions about the behavior of this observable in the resonance region.

In Sec.~\ref{sec:form} we give a brief  summary of the formalism, referring to our previous work~\cite{HillerBlin:2022ltm} for a detailed description.
The results of our computation compared with the available data are discussed in Sec.~\ref{sec:results}.
In Sec.~\ref{sec:outlook} we summarize our findings and give an outlook of these studies.

\section{Formalism}
\label{sec:form}
The formalism used in the present work follows that thoroughly described in our previous article~\cite{HillerBlin:2022ltm}. 

In terms of cross sections, the virtual photon asymmetries are given by~\cite{roberts_1990,Dharmawardane:2004yw,Melnitchouk:2005zr}
\begin{align}
\label{a1a2} 
A_1 &= \frac{\sigma_T^{1/2}-\sigma_T^{3/2}}{\sigma_T^{1/2}+\sigma_T^{3/2}}, 
\qquad A_2 = \frac{\sigma_{I}}{\sigma_T},
\end{align}
where $\sigma_I$ is the real part of the interference amplitude for virtual photons with longitudinal and transverse polarizations. 
The structure functions are then related to the virtual photon asymmetries via
\begin{subequations}
\label{eq:g1g2} 
\begin{align}
g_1 &= \frac{1}{\rho^2}\, F_1 
        \Big( A_1 + A_2\sqrt{\rho^2-1} \Big),
\\
g_2 &= \frac{1}{\rho^2}\, F_1 
        \Big( -A_1 + \frac{A_2}{\sqrt{\rho^2-1}} \Big),
\end{align}
\end{subequations}
with the kinematic factor $\rho^2 = 1 + Q^2/\nu^2$. Here, $-Q^2$ is the 4-momentum transfer squared between the electron and the proton, while $\nu$ is the virtual photon energy in the lab frame. It is related to the invariant mass $W$ of the virtual photon--target proton system via
$\nu = (W^2-M^2+Q^2)/2M$,
where $M$ is the nucleon mass.

The coherent sum of contributions from the resonances $R$ to the inclusive structure functions can be written as~\cite{Carlson:1998gf,Melnitchouk:2005zr}
\begin{subequations}
\label{Eq:coherent}
\begin{widetext}
\begin{align}
   \left(1+\frac{Q^2}{\nu^2}\right) g_1^\text{res}
   =& M^2\sum_{IJ\eta}
   \Bigg\{
     \bigg| \sum_{R^{IJ\eta}} G_+^{R^{IJ\eta}} \bigg|^2
   - \bigg| \sum_{R^{IJ\eta}} G_-^{R^{IJ\eta}} \bigg|^2
\nonumber\\
    & +\, 
    \frac{\sqrt{2Q^2}}{\nu}\,
    \Re\bigg[
    \bigg( \sum_{R^{IJ\eta}} G_0^{{R^{IJ\eta}}} \bigg)
    \bigg( \sum_{R^{IJ\eta}}(-1)^{J_{R^{IJ\eta}} - \frac12}\, \eta_{R^{IJ\eta}}
    G_+^{R^{IJ\eta}} \bigg)^\ast
    \bigg]
    \Bigg\},
\\
    \left(1+\frac{Q^2}{\nu^2}\right) g_2^\text{res}
    =& -M^2 \sum_{IJ\eta}
    \Bigg\{
      \bigg| \sum_{R^{IJ\eta}} G_+^{R^{IJ\eta}} \bigg|^2 
    - \bigg| \sum_{R^{IJ\eta}} G_-^{R^{IJ\eta}} \bigg|^2
\nonumber\\
    & -\, 
    \frac{\nu\sqrt{2}}{\sqrt{Q^2}}\,
    \Re\bigg[
    \bigg( \sum_{R^{IJ\eta}} G_0^{{R^{IJ\eta}}} \bigg) 
    \bigg( \sum_{R^{IJ\eta}}(-1)^{J_{R^{IJ\eta}}-\frac12}\, \eta_{R^{IJ\eta}}
    G_+^{R^{IJ\eta}} \bigg)^\ast
    \bigg]
    \Bigg\},
\end{align}
\end{widetext}
for the spin-dependent structure functions, and
\begin{widetext}
\begin{align}
    F_1^\text{res}
    =& M\sum_{IJ\eta}
    \Bigg\{
      \bigg| \sum_{R^{IJ\eta}} G_+^{R^{IJ\eta}} \bigg|^2
    + \bigg| \sum_{R^{IJ\eta}} G_-^{R^{IJ\eta}} \bigg|^2
    \Bigg\},
\\
    \left(1+\frac{\nu^2}{Q^2}\right) F_2^\text{res}
    =& M\nu\sum_{IJ\eta}
    \Bigg\{
      \bigg|  \sum_{R^{IJ\eta}} G_+^{R^{IJ\eta}} \bigg|^2
    + \bigg|  \sum_{R^{IJ\eta}} G_-^{R^{IJ\eta}} \bigg|^2+ 2\bigg| \sum_{R^{IJ\eta}} G_0^{R^{IJ\eta}} \bigg|^2
    \Bigg\},
\label{Eq:coherentF2}
\end{align}
\end{widetext}
\end{subequations}
for the spin-averaged structure functions.
The outer sums in Eqs.~(\ref{Eq:coherent}) run over the possible values of the spin $J$, isospin $I$, and intrinsic parity $\eta$, while the inner sums run over all resonances $R^{IJ\eta}$ that satisfy $J_R=J$, $I_R=I$ and $\eta_R=\eta$ for the spin, isospin and parity of the resonance $R$. {The amplitudes $G^{R}_{+}$, $G^{R}_{-}$, and $G^{R}_{0}$ describe the contribution from the electroexcitation amplitudes of the resonance $R$. They are related to the $\gamma^*pN^*$ electrocouplings $A_{1/2}$, $A_{3/2}$, and $S_{1/2}$ as detailed in Ref.~\cite{HillerBlin:2022ltm}. The the $\gamma^*pN^*$ electrocouplings have become available from the studies of exclusive meson electroproduction data with the CLAS detector within the mass range of $W$ $<$ 1.75 GeV and for $Q^2$ $<$ 5.0 GeV$^2$~\cite{Carman:2020qmb,Burkert:2022ioj,Blin:2019fre,Mokeev:2022xfo}.}

\section{Results}
\label{sec:results}
\begin{figure*}[th]
\begin{center}
\includegraphics[width=\textwidth]{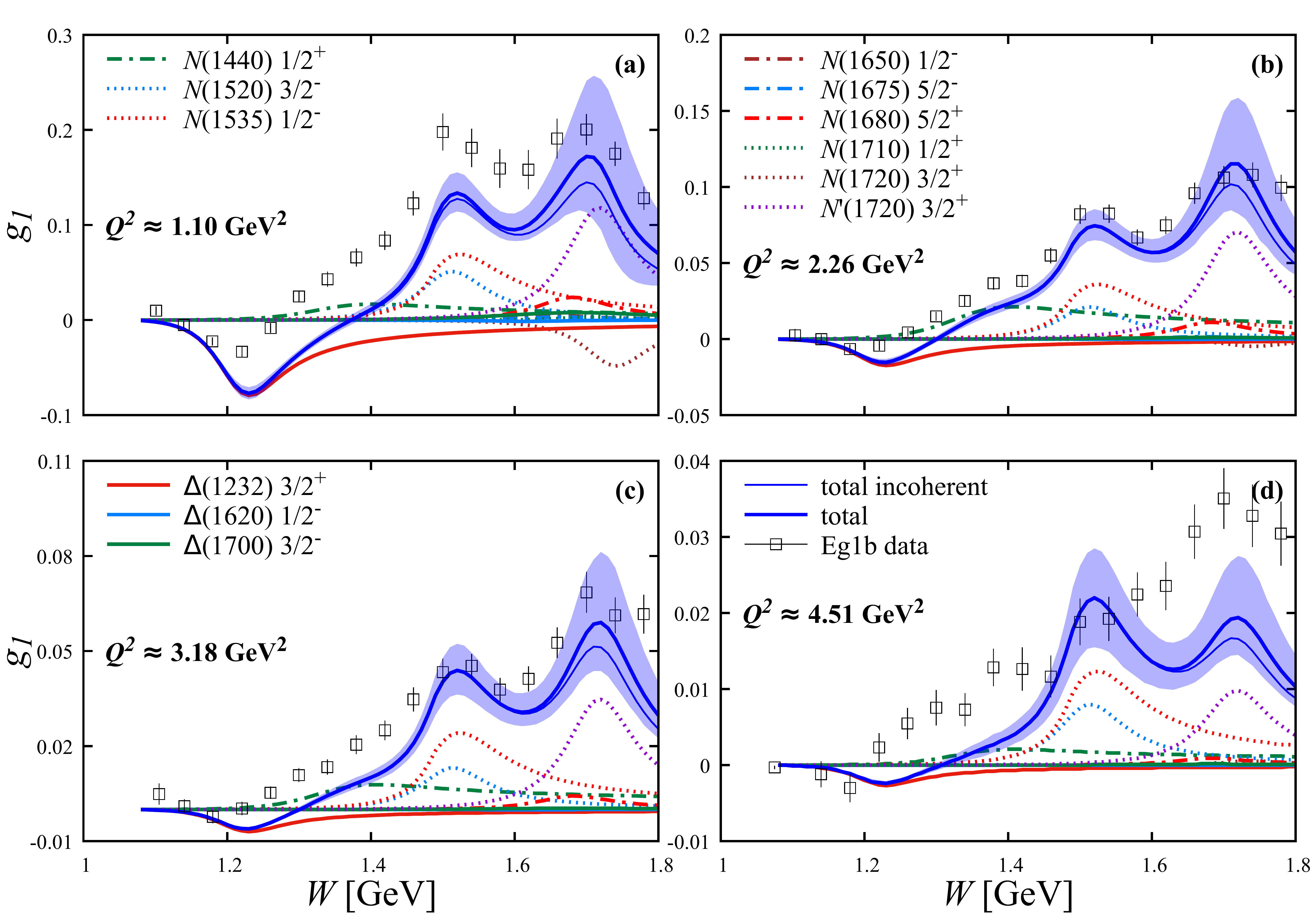}
\end{center}
\caption{Proton $g_1$ structure function data~\cite{CLAS:2017qga} (open black squares): 
{\bf (a)}~$Q^2\approx1.10$~GeV$^2$, 
{\bf (b)}~$Q^2\approx2.26$~GeV$^2$, 
{\bf (c)}~$Q^2\approx3.18$~GeV$^2$, 
{\bf (d)}~$Q^2\approx4.51$~GeV$^2$, compared to the coherent (thick blue curves) and incoherent (thin blue curves) sum of resonance contributions. 
The latter are computed at fixed $Q^2$ corresponding to the average value of the binned data in each panel. { The contributions from} individual  $N^*$ and $\Delta^*$ states 
are also shown separately. The 
uncertainties {for the resonant contributions} are computed by propagating the electrocoupling uncertainties via a bootstrap approach~\cite{Blin:2019fre}.}
\label{F:g1sing}
\end{figure*}

In Fig.~\ref{F:g1sing}, we compare the experimental results on the $g_1$ structure function measured with CLAS with {the resonant contributions, } 
{computed by employing resonance electroexcitation amplitudes deduced from exclusive CLAS electroproduction data \cite{Carman:2020qmb,Burkert:2022ioj,Blin:2019fre,Mokeev:2022xfo}. This is outlined in Sec.~\ref{sec:form}} 
~\cite{CLAS:2017qga}. We constrain ourselves to the range of $W < 1.8$~GeV and $Q^2 < 5$~GeV$^2$ where the resonance electrocouplings are currently available.  
Both the individual resonance contributions, as well as the coherent and incoherent sums of resonances are shown.

One can clearly see that the qualitative dips-and-peaks behavior in the $W$ dependence of the inclusive data is accounted for by the resonant contributions, in all $Q^2$ bins. The dominant contribution in the first resonance region is that of the $\Delta(1232)\,3/2^+$, which in turn is driven by the $G_-$ amplitude {(or by the $A_{3/2}$ electrocoupling).} According to Eq.~(\ref{Eq:coherent}), {the contribution from the $G_-$ amplitude squared enters $g_1$ with a minus sign.} This explains the negative values seen both in the $g_1$ data around that peak, as well as in the purely resonant contributions. For most of the remaining states, it is the $G_+$ amplitude, {related to the $A_{1/2}$ resonance electrocouplings,} that dominates~\cite{Blin:2019fre,HillerBlin:2022ltm}.
For this reason, the total resonant contributions to $g_1$ display a sign flip at $W$ values between the first and second resonance peaks, as is also observed in {the $W$-dependence of the measured $g_1$ 
 data~\cite{CLAS:2017qga}}, as depicted in Fig.~\ref{F:g1sing}. Our analysis confirms that the resonance contributions are the drivers of this behavior.

\begin{figure*}[th]
\begin{center}
\includegraphics[width=\textwidth]{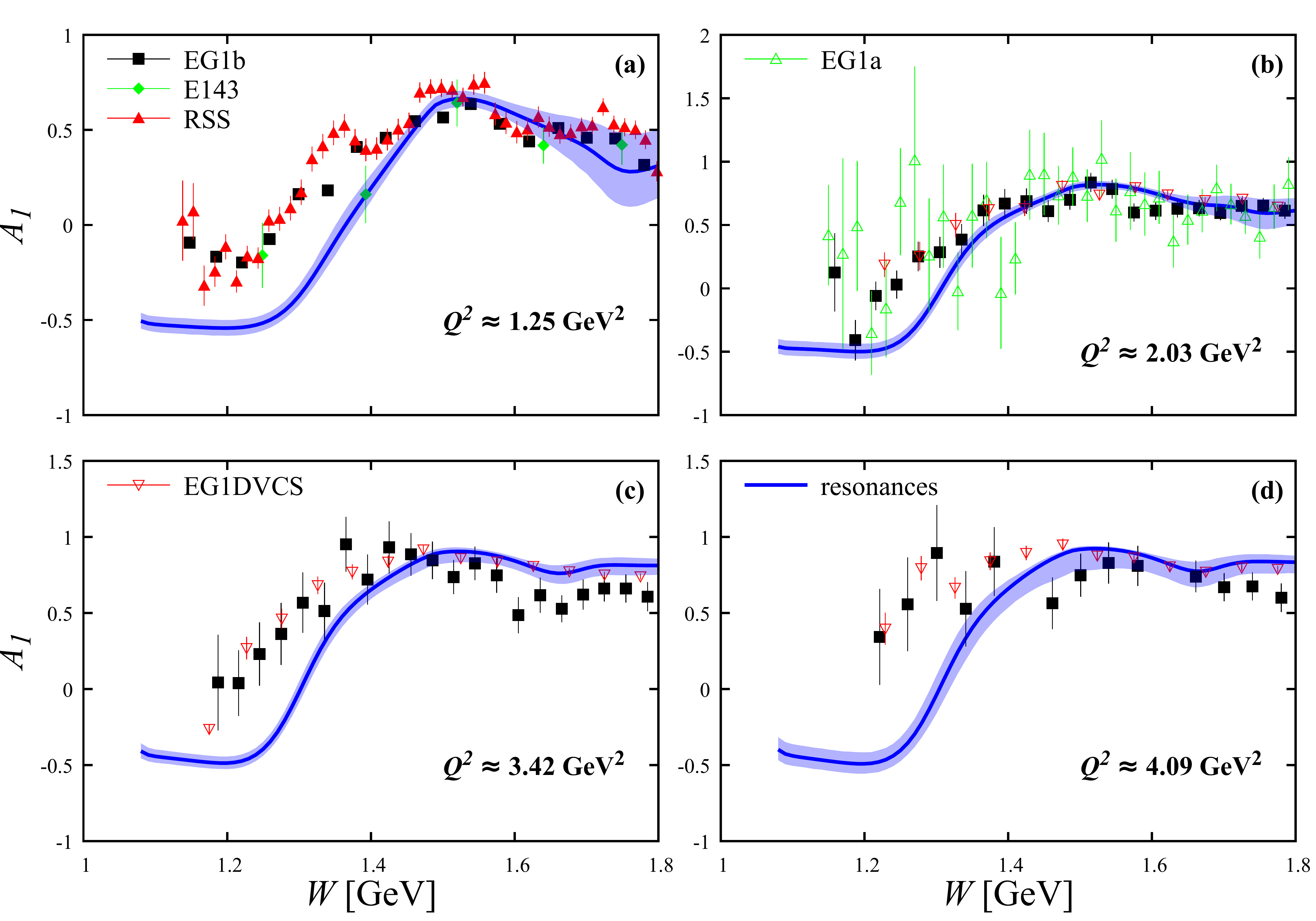}
\end{center}
\caption{ $A_{1}$ asymmetry data from the E143 experiment at SLAC~\cite{E143:1998hbs} (green filled diamonds), the RSS collaboration~\cite{RSS:2006tbm,RSS:2008ceg} (red filled triangles), as well as the CLAS experiments EG1a (green open triangles)~\cite{CLAS:2003rjt}, EG1DVCS (red open triangles)~\cite{CLAS:2014qtg}, and EG1b (black closed squares)~\cite{CLAS:2017qga}: 
{\bf (a)}~$Q^2\approx 1.25$~GeV$^2$, 
{\bf (b)}~$Q^2\approx 2.03$~GeV$^2$, 
{\bf (c)}~$Q^2\approx 3.42$~GeV$^2$, 
{\bf (d)}~$Q^2\approx 4.09$~GeV$^2$, compared to the computed purely resonant contributions (blue curves). The $1\sigma$ uncertainty bands of the resonant contributions are computed by propagating the electrocoupling uncertainties via a bootstrap approach~\cite{Blin:2019fre}.}
\label{F:A1}
\end{figure*}

In Fig.~\ref{F:A1}, we show the computed resonance contributions to the virtual photon asymmetry $A_1$, compared to {the data measured both with the  large-acceptance CLAS detector and  with other detectors of smaller acceptance in the resonance region~\cite{E143:1998hbs,CLAS:2003rjt,Bosted:2006gp,CLAS:2006ozz,RSS:2006tbm,RSS:2008ceg,CLAS:2014qtg,CLAS:2017qga}.} Since the asymmetry is defined by a cross-section ratio, the resonance structure becomes elusive. 

Nevertheless, it is intriguing to find that, {for $W$ $>$ 1.5 GeV,} 
the {$W$ and $Q^2$ evolution of $A_1$ seen in the data is} already rather well described by the inclusion of resonance contributions only. This points to {a particular sensitivity of the $A_1$ observable to the resonance contributions at $W>1.5$~ GeV. Such a behavior offers a hint for quark-hadron duality seen in this inclusive polarized electron scattering observable.} 
This finding motivates the ongoing and future studies of resonance electrocouplings with the CLAS12 detector and a possible CEBAF {energy increase up to 22 GeV~\cite{Achenbach:2023pba}}, in order to scrutinize whether this behavior holds for even larger $Q^2$ values and for the higher-mass states.

In addition, the studies presented here can and have been extended to $g_2$ and $A_2$, therefore calling for future high-acceptance measurements of these observables.

\section{Summary and outlook}

\label{sec:outlook}

In these {proceedings, we present the results on the exploration of} 
the $W$ and $Q^2$ dependence of the coherent {and incoherent} sums  of nucleon resonance contributions to the spin-dependent $g_1$ structure function and the $A_1$ virtual-photon asymmetry. These are {evaluated from the experimental results on $\gamma^*pN^*$ electrocouplings deduced from the analyses of exclusive meson electroproduction data.} As input, we used the electroexcitation amplitudes extracted from CLAS data in the mass range up to $W=1.75$~GeV ~\cite{Carman:2020qmb,Burkert:2022ioj,Blin:2019fre,Mokeev:2022xfo}.

Our findings provide evidence that the sign-flip behavior in the $g_1$ data is accounted for by the resonance contributions. In addition, the results point to {a particular sensitivity of the $A_1$ observable to the resonant contributions at $W$ $>$ 1.5 GeV.} 
This calls for further measurements at larger values of $Q^2$ and $W$, to investigate up to which QCD scales the resonant states remain sizeable and relevant. 

Further, the need to confirm the findings in this work for $g_2$ and $A_2$ gives clear motivation for future large-acceptance measurements of these observables {in experiments with polarized electron beams and for both longitudinal and transverse target polarizations}.

\begin{acknowledgements}
We thank S.~Kuhn, V.~Lagerquist, W. Melnitchouk, and P.~Pandey for useful discussions and providing us with the experimental data shown here.
This work was supported by the U.S. Department of Energy contract DE-AC05-06OR23177, under which Jefferson Science Associates, LLC operates Jefferson Lab, and by the Deutsche Forschungsgemeinschaft (DFG) through the Research Unit FOR 2926 (project number 40824754).
\end{acknowledgements}

\bibliographystyle{spphys}       
\bibliography{cite}   

\end{document}